\documentclass[twocolumn,floats,floatfix,superscriptaddress,aps,pra]{revtex4}
\usepackage{amsfonts,amssymb,amsmath}
\usepackage{color,calc}
\usepackage[dvips]{graphicx}
\usepackage{bm}

\def\be{ \begin{equation} }
\def\ee{ \end{equation} }
\def\bea{ \begin{eqnarray} }
\def\eea{ \end{eqnarray} }
\def\bse{ \begin{subequations} }
\def\ese{ \end{subequations} }
\def\i{i}
\def\e{\,\text{e}}
\def\rabi{\alpha}
\def\U{\mathbf{U}}

\def\H{\mathbf{H}}
\def\R{\mathbf{R}}

\def\c{\mathbf{c}}
\def\b{\mathbf{b}}
\def\sech{\text{sech}}

\def\half{\tfrac12}

\def\sec{\section}

\def\eps{\epsilon}
\def\IRZ{iRZ}

\begin{document}

\author{Boyan T. Torosov}
\affiliation{Institute of Solid State Physics, Bulgarian Academy of Sciences, 72 Tsarigradsko chauss\'{e}e, 1784 Sofia, Bulgaria}
\author{Nikolay V. Vitanov}
\affiliation{Department of Physics, St Kliment Ohridski University of Sofia, 5 James Bourchier blvd, 1164 Sofia, Bulgaria}

\title{Robust high-fidelity coherent control of two-state systems by detuning pulses}

\date{\today}

\begin{abstract}
Coherent control of two-state systems is traditionally achieved by resonant pulses of specific Rabi frequency and duration, by adiabatic techniques using level crossings or delayed pulses, or by sequences of pulses with precise relative phases (composite pulses).
Here we develop a method for high-fidelity coherent control which uses a sequence of \emph{detuning pulses}.
By using the detuning pulse areas as control parameters, and driving on an analogy with composite pulses, we report a great variety of detuning pulse sequences for broadband and narrowband transition probability profiles.
\end{abstract}

\maketitle


\sec{Introduction}
Coherent control is a powerful technique used in many areas of contemporary physics, e.g., in quantum optics \cite{Shore1990,QuantumOptics}, spectroscopy \cite{spectroscopy}, femtosecond \cite{femto} and attosecond \cite{atto} physics, chemical reactions \cite{chemicalReactions}, nuclear magnetic resonance \cite{NMR}, and quantum information \cite{QInfo}, to mention just a few.
The most popular techniques for coherent control include resonant pulses, adiabatic following, optimal control, and composite pulses (CPs).
These techniques use precise amplitude control (resonant pulses of precise area) \cite{Shore1990}, detuning chirping (level-crossing adiabatic following) \cite{Vitanov2001}, delayed Raman pulses [stimulated Raman adiabatic passage (STIRAP) \cite{Vitanov2017}], amplitude and detuning shaping (optimal control), and phase control (composite pulses \cite{composite}).

In this paper, we present a different method for coherent control, which uses short pulse-like variations in the detuning, while keeping the Rabi frequency constant, which we call \emph{detuning pulses}.
If these pulses are short enough and with appropriate temporal areas, they can be used as a control tool to shape the transition probability profile in essentially any desired manner.
In the limit of infinitely short time duration the detuning pulses reduce to composite pulses.
Therefore, the detuning pulses extend and generalize the concept of composite pulses and allow this concept to find applications in physical systems and processes where composite pulses are impossible or difficult to implement.

This paper is organized as follows.
The idea is introduced in Sec.~\ref{Sec:model} by first considering an exactly soluble two-state model, where a single detuning pulse or a couple of such detuning pulses are present.
The detuning pulses have a hyperbolic-secant shape and the Rabi frequency is constant.
We derive the propagator for this model and show that in a certain limit broadband or narrowband transition profiles for both partial and complete population transfer can be achieved.
The technique is then interpreted by drawing an analogy with the concept of composite pulses in the limit of very narrow (delta-function-shaped) detuning pulses in Sec.~\ref{Sec:CP}, which allows us to generalize the idea to different types and arbitrarily long detuning pulse sequences in Sec.~\ref{Sec:sequences}.
The possible implementations and benefits of the proposed techniques are discussed in Sec.~\ref{Sec:implementation}, and Sec.~\ref{Sec:summary} presents the conclusions.


\begin{figure}[tb]
	\includegraphics[width=7cm]{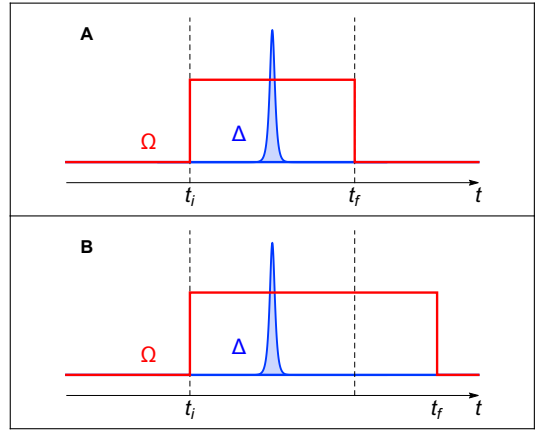}
	\caption{Schematic illustration of the inverted Rosen-Zener model used to create equal-superposition states.
		The Rabi frequency is represented by a rectangular pulse of duration $T$ (top frame) and $3T/2$ (bottom frame).
		The detuning is represented by one sech-shaped pulse centered at time $T/2$.
		The detuning pulses are intentionally scaled down compared to their actual magnitude in order to fit into the same scale as the Rabi frequency and the zero level has been shifted up for readability.
}
	\label{fig1}
\end{figure}

\begin{figure}[tb]
\includegraphics[width=\columnwidth]{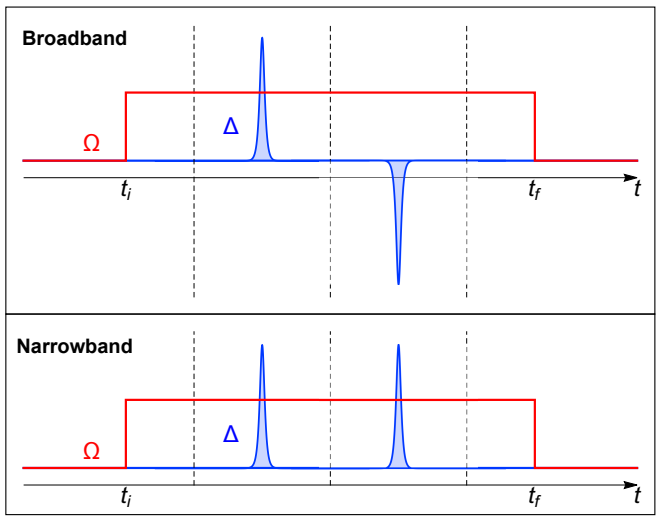}
\caption{Schematic illustration of the inverted Rosen-Zener model used to generate complete population inversion.
The Rabi frequency is represented by a rectangular pulse of duration $3T$.
The detuning is represented by two sech-shaped pulses centered at times $T$ and $2T$.
The detuning pulses are intentionally scaled down compared to their actual magnitude in order to fit into the same scale as the Rabi frequency and the zero level has been shifted up for readability.
}
\label{fig2}
\end{figure}

\sec{Analytic model}\label{Sec:model}
We consider several analytically soluble models, schematically illustrated in Figs.~\ref{fig1} and \ref{fig2}.
The two models (A and B) in Fig.~\ref{fig1} produce coherent superposition states in a broadband manner, while the two models in Fig.~\ref{fig2} produce broadband (BB) and narrowband (NB) complete population inversion profiles, respectively.

 For the cases in Fig.~\ref{fig1}, we have a rectangular resonant pulse of duration $T$ (case A) or $3T/2$ (case B), during which a sech-shaped (hyperbolic-secant) detuning pulse is applied at time $T/2$.
The Rabi frequency is $\rabi \pi/T$, where the parameter $\rabi$ is a dimensionless measure of the pulse area in units $\pi$.
The detuning pulse shape is
\be
\Delta(t) = \Delta_0 
\sech \left(\frac{t-T/2}{\tau}\right) ,
\ee
where $\Delta_0$ is a real constant and $\tau$ is the width of the detuning pulse.
There is no static detuning, i.e. in the absence of detuning pulses the driving field is on exact resonance with the transition frequency.
We will refer to this detuning pulse as the \emph{inverted  Rosen-Zener model} (\IRZ).
In the original Rosen-Zener (RZ) model \cite{RZ}, the detuning is constant and the Rabi frequency has a sech shape, while now they are reversed.
The \IRZ~and RZ models can be connected by a rotation of the basis states at angle $\pi/4$, as we will show below, and hence we can use the known solution to the RZ model to find the solution to the \IRZ~model.
As designated in Fig.~\ref{fig1}, we can view the total interaction as composed of one \IRZ~segment of length $T$ in the upper frame, or as one \IRZ~segment of length $T$ followed by one resonant pulse of duration $T/2$ in the lower frame.
We shall use this picture in order to calculate the propagator.

Similarly, for the cases in Fig.~\ref{fig2} we have a rectangular resonant pulse of duration $3T$, during which two sech-shaped detuning pulses are applied at times $T$ and $2T$.
The detunings may have opposite signs or the same sign, which, as we will show, lead to BB and NB excitation profiles, respectively.
We can view the total interaction as composed of two \IRZ~segments of length $T$, sandwiched by two resonant pulses of duration $T/2$.
We solve these models analytically by deriving the propagator of each of the segments and taking the total product.

\subsection{Single detuning pulse}

First, we consider the two models of Fig.~\ref{fig1}.
They are described by the Hamiltonian
\be\label{Hamiltonian}
\H(t) = \frac{\hbar}{2}\left[\begin{array}{cc}
	-\Delta(t) & \Omega(t) \\  \Omega(t) & \Delta(t)
\end{array}\right] ,
\ee
and the dynamics is derived by solving the Schr\"odinger equation,
\be\label{Schrodinger}
\i\hbar\partial_t\c(t) = \H(t)\c(t),
\ee
where $\c(t)=[c_1(t),c_2(t)]$ is a column-vector, containing the probability amplitudes of the two states.
The evolution of the system is described by the propagator $\U$, which maps the probability amplitudes at the initial moment $t_i$ to the final values at $t_f$: $\c(t_f)=\U(t_f,t_i)\c(t_i)$.
By applying the rotation of the basis states, $\b(t) = \R(\pi/4) \c(t)$, where
\be
\R(\theta) = \left[\begin{array}{cc}
	\cos\theta & \sin\theta \\
	-\sin\theta & \cos\theta
\end{array}\right],
\ee
we obtain the Hamiltonian in the rotated basis, $\H_{RZ}(t) = \R(\pi/4) \H(t) \R(-\pi/4)$, or explicitly,
\be\label{Hamiltonian-RZ}
\H_{RZ}(t) = \frac{\hbar}{2}\left[\begin{array}{cc}
	\Omega(t) & \Delta(t) \\  \Delta(t) & -\Omega(t)
\end{array}\right] .
\ee
This is the Hamiltonian for the exactly soluble RZ model \cite{RZ}.
The Rosen-Zener propagator $\U^{(1)}_{\text{RZ}}$ reads \cite{RZ}
\be\label{URZ}
\U^{(1)}_{\text{RZ}}=\left[\begin{array}{cc}
	a & b \\ - b^\ast & a^\ast
\end{array}\right],
\ee
where
\footnote{The factor $\exp(-\i\rabi\pi/2)$ comes from using the symmetric Schr\"{o}dinger representation \eqref{Hamiltonian} of the Hamiltonian.}
\bse\label{ab-RZ}
\begin{align}
&a=\frac{\Gamma^2\left[ \half(1-\i\rabi\tau) \right]\exp(-\i\rabi\pi/2)}{\Gamma\left[ \half(1-\i\rabi\tau -\Delta_0\tau) \right]\Gamma\left[ \half(1-\i\rabi\tau +\Delta_0\tau) \right]} ,\\
&b=-\i\frac{\sin(\half\pi\Delta_0\tau)}{\cosh(\half\pi\rabi\tau)} .
\end{align}
\ese
This solution can be used to derive the solution for the original Hamiltonian \eqref{Hamiltonian}.
The total propagators for the two models of Fig.~\ref{fig1} are
\bse
\begin{align}
&\U^{(A)} = \R(-\pi/4) \U^{(1)}_{\text{RZ}} \R(\pi/4), \\
&\U^{(B)} = \U_{\pi/2}\R(-\pi/4) \U^{(1)}_{\text{RZ}} \R(\pi/4).
\end{align}
\ese
Here
\be
\U_{\pi/2} = \left[
\begin{array}{cc}
	\cos\left( \pi\rabi/4 \right) & -\i\sin\left( \pi\rabi/4 \right) \\
	-\i\sin\left( \pi\rabi/4 \right) & \cos\left( \pi\rabi/4 \right)
\end{array}
\right]
\ee
is the propagator of a resonant pulse of area $\pi\rabi/2$.
For $\alpha=1$ this is a nominal (i.e. in the absence of errors) $\pi/2$ pulse.

\begin{figure}[tb]
	\includegraphics[width=0.8\columnwidth]{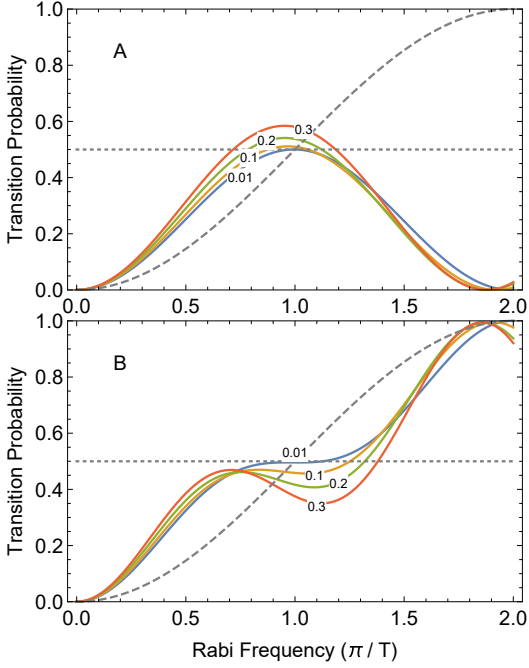}
	\caption{ Transition probability as a function of the Rabi frequency for the \IRZ~model in the superposition cases A and B of Fig.~\ref{fig1}, for detuning pulse width $\tau=0.01, 0.1 ,0.2 ,0.3$ (in units $T/\pi$).
For comparison, the dashed line shows the profile of a single resonant pulse, $\sin^2(\pi\rabi/4)$ \cite{note-half}.
	}
	\label{figSup}
\end{figure}

We are now interested in the case when the width of the detuning pulse is small ($\tau\ll T$).
We take the limit $\tau\to 0$ by keeping the detuning pulse area $\pi\Delta_0 \tau$ fixed, and therefore we have $\Delta_0\to\infty$.
%
The transition probability $P=\left| \U_{12} \right|^2$ in the limit of $\tau\to 0$ reads
\bse
\begin{align}
&P^{(A)} = \cos^2\frac{\pi\delta}{2} \sin^2\frac{\pi\rabi}{2}, \\
&P^{(B)} = \sin^2\frac{\pi\delta}{2} \sin^2\frac{\pi\rabi}{4} + \cos^2\frac{\pi\delta}{2} \sin^2\frac{3\pi\rabi}{4},
\end{align}
\ese
where we have introduced the dimensionless parameter $\delta=\Delta_0\tau$.
If we expand near $\alpha=1$ we find
\bse
\begin{align}
P^{(A)} =& \cos^2\frac{\pi\delta}{2} \left[ 1 -\frac{\pi^2}{4} \eps^2  +O(\eps^4)\right], \\
P^{(B)} =& \frac{1}{2} - \frac{\pi}{4}(1+2\cos\pi\delta)\eps + O(\eps^3),
\end{align}
\ese
with $\rabi=1+\eps$.
Now, if we set $\delta=\frac12$ for case A, and $\delta=\frac23$ for case B, we obtain, respectively,
\bse
\begin{align}
P^{(A)} =& \frac{1}{2} + O(\eps^2), \\
P^{(B)} =& \frac{1}{2} + O(\eps^3).
\end{align}
\ese
In other words, an equal superposition state is generated in each case, with accuracy up to the second order of the error $\eps$ in case A and up to the third order in case B.
In either cases, this is a dramatic improvement in the resilience to pulse area errors over the case of exact resonance when $P = \sin^2(\pi\alpha/4)$, which is accurate only to the first order in the pulse area error at $\rabi=1$ \cite{note-half}, 
\be
P^{(\text{res})} = \frac{1}{2} + O(\eps),
\ee
with $\rabi=1+\eps$.

The excitation profiles generated by the two \IRZ models A and B are shown in Fig.~\ref{figSup}.
The quadratic and qubic stabilization of the transition probability at the value $\frac12$ is clearly visible for a sufficiently narrow detuning pulse.
As the detuning pulse width $\tau$ increases the transition probability begins to deviate from the desired value of $\frac12$.

\subsection{Two detuning pulses: broadband profiles}

\begin{figure}[tb]
	\includegraphics[width=0.8\columnwidth]{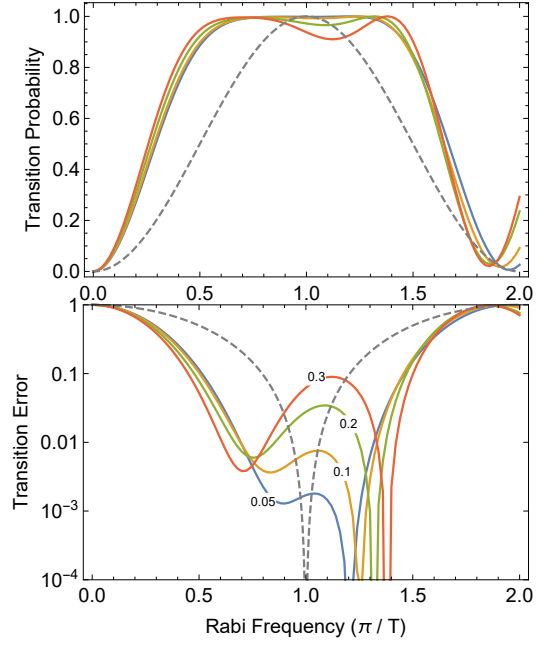}
	\caption{(Top frame) Transition probability as a function of the Rabi frequency for the composite \IRZ~model in the BB case, for detuning pulse width $\tau=0.05, 0.1, 0.2, 0.3$ (in units $T/\pi$). For comparison, the dashed line shows the $\sin^2$ profile of a single resonant pulse.
(Bottom frame) Corresponding transition probability error in logarithmic scale.
	}
	\label{figBB}
\end{figure}

For the two cases of Fig.~\ref{fig2} there are two detuning pulses and hence we refer to them as the composite \IRZ model.
Consequently, the propagators are more involved than for a single detuning pulse.
For the BB case, illustrated in Fig.~\ref{fig2} (top), the total propagator is
\be\label{UBB}
\U^{\text{BB}} = \U_{\pi/2}\R(-\pi/4)\U^{(2)}_{\text{RZ}} \U^{(1)}_{\text{RZ}} \R(\pi/4)\U_{\pi/2},
\ee
where the RZ propagator  $\U^{(1)}_{\text{RZ}}$ is given by Eq.~\eqref{URZ}, and the second RZ propagator  $\U^{(2)}_{\text{RZ}}$ is
\be
\U^{(2)}_{\text{RZ}}=\left[
\begin{array}{cc}
a & -b \\
 b^\ast & a^\ast
\end{array}
\right].
\ee
The explicit expression of Eq.~\eqref{UBB} is far too cumbersome to be presented here.
In the limit of small $\tau$, we expand $\U^{\text{BB}}_{11}$ near $\rabi=1$,
\footnote{The reason to expand $\U^{\text{BB}}_{11}$ and not $\U^{\text{BB}}_{12}$ is just for derivation simplicity.}
\be
\U^{\text{BB}}_{11} = \frac{\pi}{2}\left[1+2\cos\pi\delta\right]\eps + O(\eps^3) ,
\ee
with $\rabi = 1 + \eps$.
For $\delta = \frac23$ the leading term is canceled, and $\U^{\text{BB}}_{11}$ becomes of order $O(\eps^3)$.
Hence the transition probability $|\U^{\text{BB}}_{21}|^2 = 1 - |\U^{\text{BB}}_{11}|^2$ is of order $1 - O(\eps^6)$.
Therefore, the excitation profile is far broader than the one for a single resonant pulse which is of order $1 - O(\eps^2)$.

The BB profiles for different values of $\tau$ are depicted in Fig.~\ref{figBB}, where we plot the transition probability as a function of the Rabi frequency.
As seen from the figure the composite \IRZ~model produces BB population inversion profiles, which are far broader than the one for a single resonant pulse.
They become more accurate as the detuning pulse width decreases.

\subsection{Two detuning pulses: narrowband profiles}

\begin{figure}[tb]
	\includegraphics[width=0.8\columnwidth]{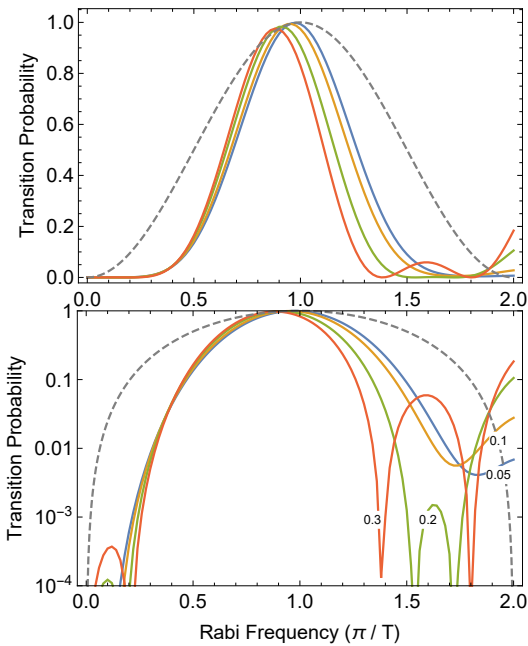}
	\caption{(Top frame) Transition probability as a function of the Rabi frequency for the composite \IRZ~model in the NB case, for detuning pulse width $\tau=0.05, 0.1 ,0.2 ,0.3$ (in units $T/\pi$). For comparison, the dashed line shows the $\sin^2$ profile of a single resonant pulse. (Bottom frame) Transition probability in logarithmic scale.
	}
	\label{figNB}
\end{figure}

In a similar manner, for the NB case, illustrated in Fig.~\ref{fig2} (bottom), the total propagator is
\be
\U^{\text{NB}} = \U_{\pi/2}\R(-\pi/4)\U^{(1)}_{\text{RZ}} \U^{(1)}_{\text{RZ}} \R(\pi/4)\U_{\pi/2}.
\ee
In order to produce an NB profile, now we expand $\U^{\text{NB}}_{12}$ (rather than $\U^{\text{NB}}_{11}$) at $\rabi=2$,
\be\label{Unb}
\U^{\text{NB}}_{12} = \i\frac{\pi}{2}\left[1+2\cos(\pi\delta)\right](\eps) + O(\eps^3),
\ee
with $\rabi = 2 + \eps$.
One can also easily check that
\be
\U^{\text{NB}}_{12}|_{\tau=0,\rabi=1} = \i,
\ee
which corresponds to complete population inversion at $\rabi=1$.
As in the BB case, we set $\delta = 2/3$ in order to cancel the leading term in \eqref{Unb}, and hence we achieve
$\U^{\text{NB}}_{12} = O(\eps^3)$, and hence transition probability $|\U^{\text{NB}}_{12}|^2 = O(\eps^6)$ in the vicinity of $\rabi = 2$.
Therefore the low-probability range around $\rabi = 2$ gets broader compared to the one for a single resonant pulse, which is $O(\eps^2)$,
 and consequently, the high-probability range around $\rabi=1$ is squeezed compared to a single pulse.

The profiles for the NB case are illustrated in Fig.~\ref{figNB}, where we plot the transition probability vs the Rabi frequency. As seen from the figure, the composite \IRZ~model can be used to produce NB profiles, given that the width of the detuning pulses is small enough. Figs.~\ref{figBB} and \ref{figNB} demonstrate that the model starts to fail to produce the desired profiles at $\tau \gtrsim 0.1 T/\pi$, i.e. when the width of the detuning pulses exceeds about 3\% of the separation between them.


\section{Detuning pulses and composite pulses}\label{Sec:CP}

\subsection{Composite pulses}

In this section we will briefly introduce the concept of composite pulses and we will show how the behavior of our exactly soluble models can be explained within this framework.
The composite pulses are used in the context of the dynamics of a two-state quantum system, driven by an external coherent field.
Such a system is described by the Schr\"{o}dinger equation \eqref{Schrodinger}.
The propagator $\U$ is expressed by Eq.~\eqref{URZ}.
However, now the Cayley-Klein parameters $a$ and $b$ are no longer given by the RZ formulas \eqref{ab-RZ} but are arbitrary complex numbers restricted by the condition $|a|^2+|b|^2 = 1$.
The composite pulses are derived by introducing a constant phase shift $\phi$ in the driving field, $\Omega(t)\to\Omega(t)\e^{\i\phi}$, which is mapped onto the propagator as
 \be
\U_{\phi}=\left[\begin{array}{cc}
a & b \e^{\i\phi}\\
 -b^{\ast}\e^{-\i\phi} & a^{\ast}
\end{array}\right] .
\ee
Instead of a single pulse, we take a sequence of $N$ pulses, each with a phase $\phi_k$.
The total propagator of such a composite sequence is
\be
\U^{N} = \U_{\phi_N}\U_{\phi_{N-1}}\dots\U_{\phi_2}\U_{\phi_1},
\ee
where the composite phases $\phi_k$ are used as control parameters to shape the excitation profile in a desired manner.
In such a way one can produce a huge variety of CPs, which are well documented in the literature.

\subsection{Relation of detuning pulses to composite pulses}

We shall now elucidate the relation of our composite \IRZ~model to composite pulses.
We begin with the interaction representation of the Schr\"{o}dinger equation, which is derived after a simple phase transformation of the amplitudes
\be
\c(t)\to
\left[
\begin{array}{cc}
	\e^{\i D(t)/2} & 0 \\
	0 & \e^{-\i D(t)/2}
\end{array}
\right]
\c(t) ,
\ee
where $D(t)=\int_{t_i}^{t}\Delta(t^{\prime})d t^{\prime}$ . In this representation, the Hamiltonian changes its form into
\be
\H(t) = \frac{\hbar}{2}\left[\begin{array}{cc}
0 & \Omega(t)\e^{-\i D(t)} \\
 \Omega(t) \e^{\i D(t)} & 0
\end{array}\right] .
\ee
Hence the integral of the detuning can be seen as a phase shift in the Rabi frequency, i.e. the phase shift is equal to the temporal area of the respective detuning pulse.
In the limit of a detuning pulse shaped as a Dirac delta function, the integral becomes a Heaviside step function, which is exactly a phase jump as in a composite pulse.
However, for a detuning pulse of a finite duration the ensuing phase change has a finite duration too.
Hence, instead of using a composite sequence of pulses with phase jumps in the driving field Rabi frequency we can just shape the detuning in the form of a sequence of pulses with specific areas.
Therefore, the detuning pulses generalize the concept of composite pulses to phase jumps of finite duration and reduce to them in the limit of infinitely short detuning pulse duration.

Now we can go back to the composite \IRZ~model of Fig.~\ref{fig2} and see it from this new perspective.
Instead of a couple of $\pi$ pulses with $\sech$ detunings, surrounded by a couple of $\pi/2$ resonant pulses, we can look at it as three pulses of duration $T$ with two $\sech$-shaped detunings pulses in-between.
In the limit of very short width of the detuning pulses, this maps to phase jumps between the three pulses, and hence we have a three-component composite pulse.
For both BB and NB cases, we found that the optimal value of $\delta$ is $\frac23$.
Therefore, we obtained generalizations of composite sequences with relative phases $\phi_k=(0,\frac23\pi,0)$ for the BB case and $\phi_k=(0, \frac23\pi, -\frac23\pi)$ for the NB case, which are exactly the phases, derived earlier \cite{TorosovBB,VitanovNB}.

\begin{figure}[tb]
	\includegraphics[width=8.5cm]{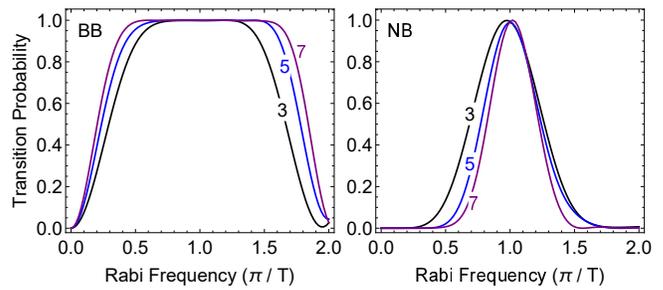}
	\caption{Transition probability as a function of the Rabi frequency for BB and NB detuning pulse sequences with 3, 5 and 7 pulses and $\tau=0.05 T/\pi$.
}
	\label{figBbNb}
\end{figure}

\section{Detuning pulse sequences}\label{Sec:sequences}

\subsection{Broadband and narrowband  pulses}

\begin{figure}[tb]
\includegraphics[width=0.8\columnwidth]{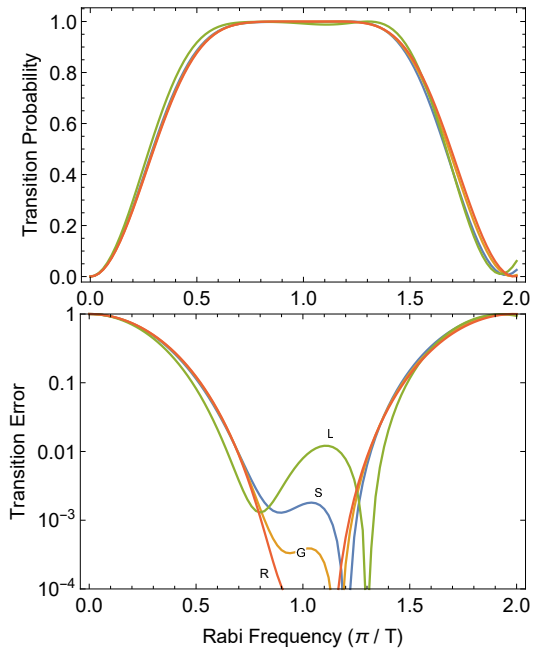}
\caption{Transition probability (top frame) and transition probability error (bottom frame) as a function the Rabi frequency for BB sequences of 3 detuning pulses with different shapes: hyperbolic secant (S), gaussian (G), lorentzian (L), and rectangular (R).}
\label{figDiffShape}
\end{figure}

We will now show how to apply our technique on the standard broadband (BB) and narrowband (NB) pulses, which expand or squeeze the high-fidelity transition probability profile compared to a single $\pi$ pulse.
We will use the CPs derived in \cite{TorosovBB} and \cite{VitanovNB}.
The BB  CPs consist of symmetric sequences of $N$ identical resonant pulses (where $N=2n+1$ is an odd number) with phases
\be
(\phi_1,\phi_2,\ldots,\phi_n,\phi_{n+1},\phi_n,\ldots,\phi_2,\phi_1),
\ee
 where $\phi_k=k(k-1)n\pi/N$  ($k=1,\ldots ,n+1$) \cite{TorosovBB}.
The NB CPs also consist of an odd number of identical pulses $N=2n+1$, but with different phases,
\be
(0,\phi_1,-\phi_1,\phi_2,-\phi_2,\ldots,\phi_n,-\phi_n),
\ee
 where $\phi_k=2k\pi/N$  ($k=1,\ldots ,n$) \cite{VitanovNB}.
Using our detuning pulses we can achieve the same result by just setting a constant Rabi frequency and introducing detuning pulses with areas equal to the respective phase jump $\int\Delta_k(t)d t = \phi_{k+1}-\phi_{k}$  ($k=1,\ldots ,N$) (the integration is over the duration of the detuning pulse).
The three-pulse BB and NB sequences are the cases schematically illustrated in Fig.~\ref{fig2}.
The exact shape of these detuning pulses is of no importance, but the smaller is their duration, the better is the approximation to the standard CPs.
More generally, we can produce a $N$-pulse composite sequence (with $N$ odd) by imposing a pulse with constant Rabi frequency $\Omega$ and duration $N T$, where
\bse
\begin{align}\label{BBN}
  & \Omega T = \rabi\pi, \\
  & \Delta(t) = \Delta_0 \sum_{k=1}^{N-1}(\phi_{k+1}-\phi_{k}) f(t-kT) , \label{BB-Delta}
\end{align}
\ese
where $f(t)$ describes the pulse shape and is normalized such that $\int f(t) dt = 1/\Delta_0$.

In Fig.~\ref{figBbNb} we plot the transition probability as a function of the Rabi frequency for sequences of 3, 5 and 7 detuning pulses.
Obviously, our method produces the expected BB and NB excitation profiles, which are quite similar to the original CP profiles.
As we already noted, the particular shape of the detuning pulses has no significance as far as the condition of short pulse width is fulfilled.
This is demonstrated in Fig.~\ref{figDiffShape}, where four different pulse shapes are compared.

\begin{figure}[tb]
\includegraphics[width=8.5cm]{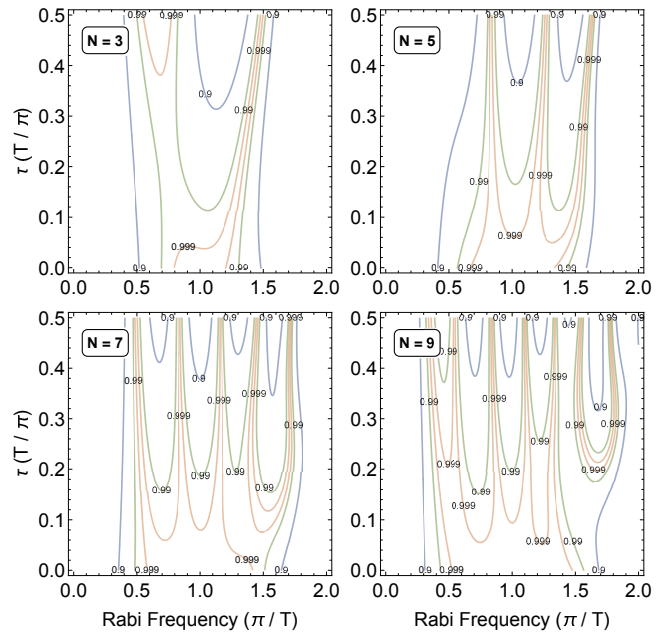}
\caption{Transition probability as a function of the Rabi frequency and the detuning pulse width for BB composite sequences with 3, 5, 7 and 9 pulses.
}
\label{figVsT}
\end{figure}


\begin{figure}[tb]
\includegraphics[width=8.5cm]{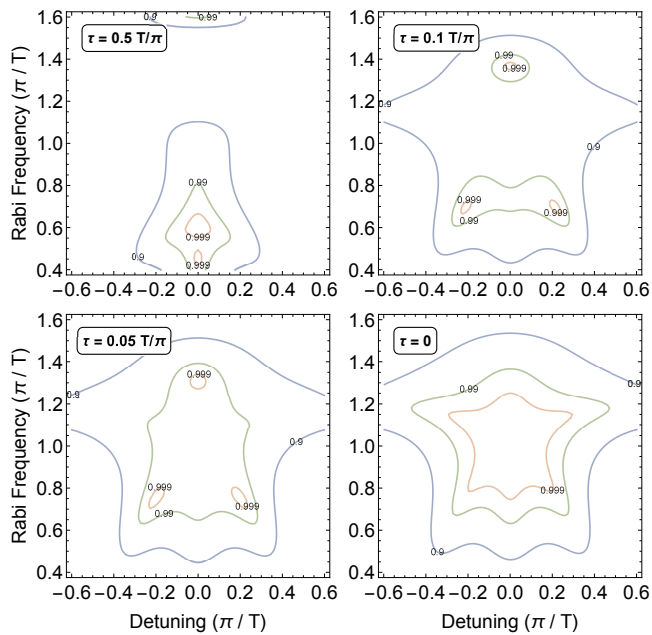}
\caption{Transition probability as a function of the detuning and the Rabi frequency for a universal composite sequence with $N=5$ and different value for the detuning pulse duration. The bottom right frame corresponds to the original universal composite sequence, derived in \cite{GenovUniversal}.}
\label{figUni}
\end{figure}

In Fig.~\ref{figVsT} we illustrate how the time duration of the detuning pulses affects the fidelity of the method. We see that for the CPs under consideration our method can be applied if $\tau\lesssim 0.1 T/\pi$. Of course, this condition depends on the particular CPs and on the fidelity that is sought, and should not be treated as some strict requirement.

We point out that by using similar argumentation one can produce analogues of the recently developed twin composite pulses \cite{twin}.
Moreover, one can produce detuning pulse sequences which deliver passband excitation profiles using the available passband composite pulses \cite{Kyoseva2013}.
Similarly, one can use the available composite $\theta$ pulses \cite{NMR} to derive detuning pulse sequences, which produce robust partial rotations of the Bloch vector at some fractional values of $\pi$ (at an angle $\theta$), e.g. to generate coherent superpositions of quantum states.

\subsection{Universal composite pulses}

The method described in this paper can turn essentially any composite pulse into a sequence of detuning pulses while largely preserving the properties of the composite pulse.
In this section we apply it to the universal composite pulses \cite{GenovUniversal}, which can be used to cancel systematic errors vs \emph{any} experimental parameter, hence the term \emph{universal}.
Our method can be easily applied on such pulses, again by just replacing the phase jumps by detuning pulses.
In the case of 5 detuning pulses, the values of the composite phases are $\phi_k=(0,5,2,5,0)\pi/6$.
As in the previous section, we consider constant Rabi frequency and hyperbolic-secant shape of the detuning pulses, but this time we add a constant error shift in the detuning.
In Fig.~\ref{figUni} we plot the transition probability for such universal composite sequence, where the four frames correspond to different values of $\tau$.
We notice that by choosing a smaller value of $\tau$ the transition profile approaches the one of the original universal composite sequence (bottom right frame).
Nevertheless, the detuning pulse sequence produces robust ranges of high transition probability even for finite detuning pulse widths.

\section{Physical relevance and implementation}\label{Sec:implementation}

The method of detuning pulse sequences can be implemented in any physical system which allows for a fast and accurate control of the detuning.
In most of the physical realizations of a qubit, there is a good control of the detuning, while a phase jump of the interaction is sometimes difficult to implement or inaccurate.
For radiofrequency or microwave fields phase jumps are easy to implement directly in the generator.
However, already for laser fields of nanosecond duration phase jumps may suffer from transient effects in the phase and the amplitude of the pulse when switching the laser field on/off by e.g., an acoustooptic modulator (AOM).
Moreover, even if the AOM produces accurate phase jumps it is always beneficial to have an alternative tool in the lab for merely practical reasons:
 detuning pulses are easy to produce because far-off-resonant laser light, which can produce Stark-shift-induced detuning jumps (as in Stark-chirp-induced rapid adiabatic passage \cite{Rickes2000,Yatsenko2002,Rangelov2005,Oberst2008}), is always available in the lab as a part of frequency conversion setups.
Moreover, detuning jumps can be produced also by pulsed magnetic or electric fields.

In addition to these examples there are physical platforms and processes wherein phase jumps are merely impossible to implement, e.g. for applications of composite sequences in waveguide and nonlinear optics.
In waveguides \cite{Ciret2013}, there is simply no way known to inflict phase changes in the coupled-mode equations that describe the propagation of light in the waveguides.
However, the propagation constants of a pair of two waveguides, the difference between which plays the role of the detuning, can be spatially controlled \cite{Oukraou2017}, thereby making it possible to implement detuning pulses and hence to achieve the great control accuracy and robustness similar to composite pulses.

Another example is nonlinear frequency conversion wherein arbitrary phase jumps in the coupling (e.g., represented by the optical nonlinearity) are impossible to implement.
It is true that one can use sign flips in the $\chi^{(2)}$ nonlinearity (phase jumps of $\pi$) \cite{Rangelov2014} and implement analogues of existing composite pulses that only require phase steps of $\pi$ \cite{Shaka1985,Shaka1987}; this, however, greatly reduces the pool of available composite sequences.
Instead one can use the detuning pulse sequences presented here to realize greater and broadband conversion efficiency \cite{Genov2014}.


\section{Summary}\label{Sec:summary}

In the present work, we presented a method for robust control of the excitation profile of a two-state quantum system by a sequence of detuning pulses.
In the limit of very short pulse duration the sequence resembles a composite pulse.
By using this analogy, we have presented a large variety of detuning pulse sequences, with which one can produce broadband, narrowband, and passband profiles for complete population transfer, and even robust coherent superpositions of states.

Composite pulses are an extremely powerful method for quantum control.
However, they suffer from an intrinsic limitation: one must be able to control the relative phases between the pulses.
In some physical systems, however, such control is not possible, or not easy.
By using the method developed in this paper, we can overcome these limitations, and exploit the full potential of composite pulses in almost any physical situation.
Therefore, the quantum control method presented here can be viewed as an alternative to the standard CPs, and can be applied whenever detuning shaping is easier to implement than phase jumps in the field.

\acknowledgments
This work was supported by the Bulgarian Science Fund Grant No. DN 18/14.



\begin{thebibliography}{99}
\bibitem{note-half}
Note that for a single pulse, $\rabi=1$ corresponds to a pulse area of $\pi/2$.
\end{thebibliography}
\end{document}